\documentclass[10pt,twocolumn,superscriptaddress,aps,prl,balancelastpage,longbibliography]{revtex4}
\usepackage[latin9]{inputenc}
\usepackage{color,multirow}
\usepackage{verbatim}
\usepackage{amsmath}
\usepackage{mathtools}
\usepackage{amssymb,ulem,manfnt}
\usepackage{graphicx,cellspace,booktabs}
\usepackage{esint}
\usepackage[unicode=true,pdfusetitle,
 bookmarks=true,bookmarksnumbered=false,bookmarksopen=false,
 breaklinks=false,pdfborder={0 0 1},backref=false,colorlinks=true]{hyperref}
\usepackage{breakurl}
\usepackage{epstopdf}
\usepackage{yfonts}
\usepackage[dvipsnames]{xcolor}

\makeatletter

\@ifundefined{textcolor}{}
{%
 \definecolor{BLACK}{gray}{0}
 \definecolor{WHITE}{gray}{1}
 \definecolor{RED}{rgb}{1,0,0}
 \definecolor{GREEN}{rgb}{0,1,0}
 \definecolor{BLUE}{rgb}{0,0,1}
 \definecolor{CYAN}{cmyk}{1,0,0,0}
 \definecolor{MAGENTA}{cmyk}{0,1,0,0}
 \definecolor{YELLOW}{cmyk}{0,0,1,0}
}

\usepackage[caption=false]{subfig}
\usepackage{bm}

\newcommand{\bra}[1]{\langle #1 |}
\newcommand{\ket}[1]{|#1\rangle}

\def\l@subsubsection#1#2{}
\makeatother
\begin{document}
\title{Many-body Chern number from statistical correlations of randomized measurements}

\author{Ze-Pei Cian}
\affiliation{Joint Quantum Institute, College Park, 20742 MD, USA}
\affiliation{The Institute for Research in Electronics and Applied Physics, University of Maryland, College Park, 20742 MD, USA} 

\author{Hossein Dehghani}
\affiliation{Joint Quantum Institute, College Park, 20742 MD, USA}
\affiliation{The Institute for Research in Electronics and Applied Physics, University of Maryland, College Park, 20742 MD, USA}

\author{Andreas Elben}
\affiliation{Center for Quantum Physics, University of Innsbruck, Innsbruck A-6020, Austria.}
\affiliation{Institute for Quantum Optics and Quantum Information of the Austrian Academy of Sciences, Innsbruck A-6020, Austria.
}

\author{Beno\^{i}t Vermersch}
\affiliation{Center for Quantum Physics, University of Innsbruck, Innsbruck A-6020, Austria.}
\affiliation{Institute for Quantum Optics and Quantum Information of the Austrian Academy of Sciences, Innsbruck A-6020, Austria.
}
\affiliation{Univ. Grenoble Alpes, CNRS, LPMMC, 38000 Grenoble, France.}

\author{Guanyu Zhu}
\affiliation{IBM T.J. Watson Research Center, Yorktown Heights, New York 10598, USA.}

\author{Maissam Barkeshli}
\affiliation{Joint Quantum Institute, College Park, 20742 MD, USA}
\affiliation{Condensed Matter Theory Center, Department of Physics, University of Maryland, College Park, 20742 MD, USA}

\author{Peter Zoller}
\affiliation{Center for Quantum Physics, University of Innsbruck, Innsbruck A-6020, Austria.}
\affiliation{Institute for Quantum Optics and Quantum Information of the Austrian Academy of Sciences, Innsbruck A-6020, Austria.
}

\author{Mohammad Hafezi}
\affiliation{Joint Quantum Institute, College Park, 20742 MD, USA}
\affiliation{The Institute for Research in Electronics and Applied Physics, University of Maryland, College Park, 20742 MD, USA}

\date{\today}
\begin{abstract}
One of the main topological invariants that characterizes several topologically-ordered phases is the many-body Chern number (MBCN). Paradigmatic examples include several fractional quantum Hall phases, which are expected to be realized in different atomic and photonic quantum platforms in the near future. Experimental measurement and numerical computation of this invariant are conventionally based on the linear-response techniques which require having access to a family of states, as a function of an external parameter, which is not suitable for many quantum simulators. Here, we propose an ancilla-free experimental scheme for the measurement of this invariant, without requiring any knowledge of the Hamiltonian. Specifically, we use the statistical correlations of randomized measurements to infer the MBCN of a wavefunction. Remarkably, our results apply to disk-like geometries that are more amenable to current quantum simulator architectures.
\end{abstract}

\maketitle

{\it Introduction.---} Topologically ordered systems are a class of gapped quantum phases of matter  \cite{wen1990topological, RevModPhys.89.041004}, which can have robust topological ground-state degeneracy, and host excited states with fractional statistics, known as anyons \cite{Wilczek1982Quantum}. These systems, unlike symmetry protected topological (SPT) phases that have short range entanglement, acquire long-range entanglement which makes them a suitable platform for realizing quantum computation \cite{KITAEV20032,RevModPhys.80.1083}. Paradigmatic examples of chiral topologically ordered systems are the fractional quantum Hall (FQH) states that in certain cases are characterized by the many-body Chern number (MBCN), as their topological invariant. 

In recent years, the interest in engineering topological states of matter in synthetic quantum systems has substantially grown. Examples of such quantum simulators include neutral atoms \cite{Cooper2019Topological}, superconducting qubits \cite{houck2012chip,roushan2017chiral}, photons \cite{ozawa2019topological}, and more recently Rydberg atoms  \cite{clark2019observation,browaeys2020many}. With these developments, the benefit of having direct access to the wave function in quantum simulators opens new avenues to investigate and measure the topological properties. In the conventional condensed matter physics the detection of topological properties relies on the application of external probes and linear response framework, and similar schemes have been also proposed for the simulated matter \cite{repellin2019detecting, tran2017probing, asteria2019measuring, repellin2020hall,motruk2020detecting}. Moreover, ancilla-based approaches have been proposed that involve a many-body Ramsey interferometry to measure the topological charge \cite{grusdt2016interferometric}, and entanglement spectrum \cite{pichler2016measurement}. But the fact that the ancilla should be coupled to the entire system limits the applicability of such schemes.  Recently,  this question has been theoretically investigated in the context of SPT systems \cite{denNijs1989preroughening, Pollmann2012symmetry, Cirac2012order, Shiozaki2017, Shiozaki2018many,elben2020many}, but the problem for topologically-ordered system has been relatively unexplored.

 \begin{figure}[b]
    \includegraphics[width=0.40\textwidth]{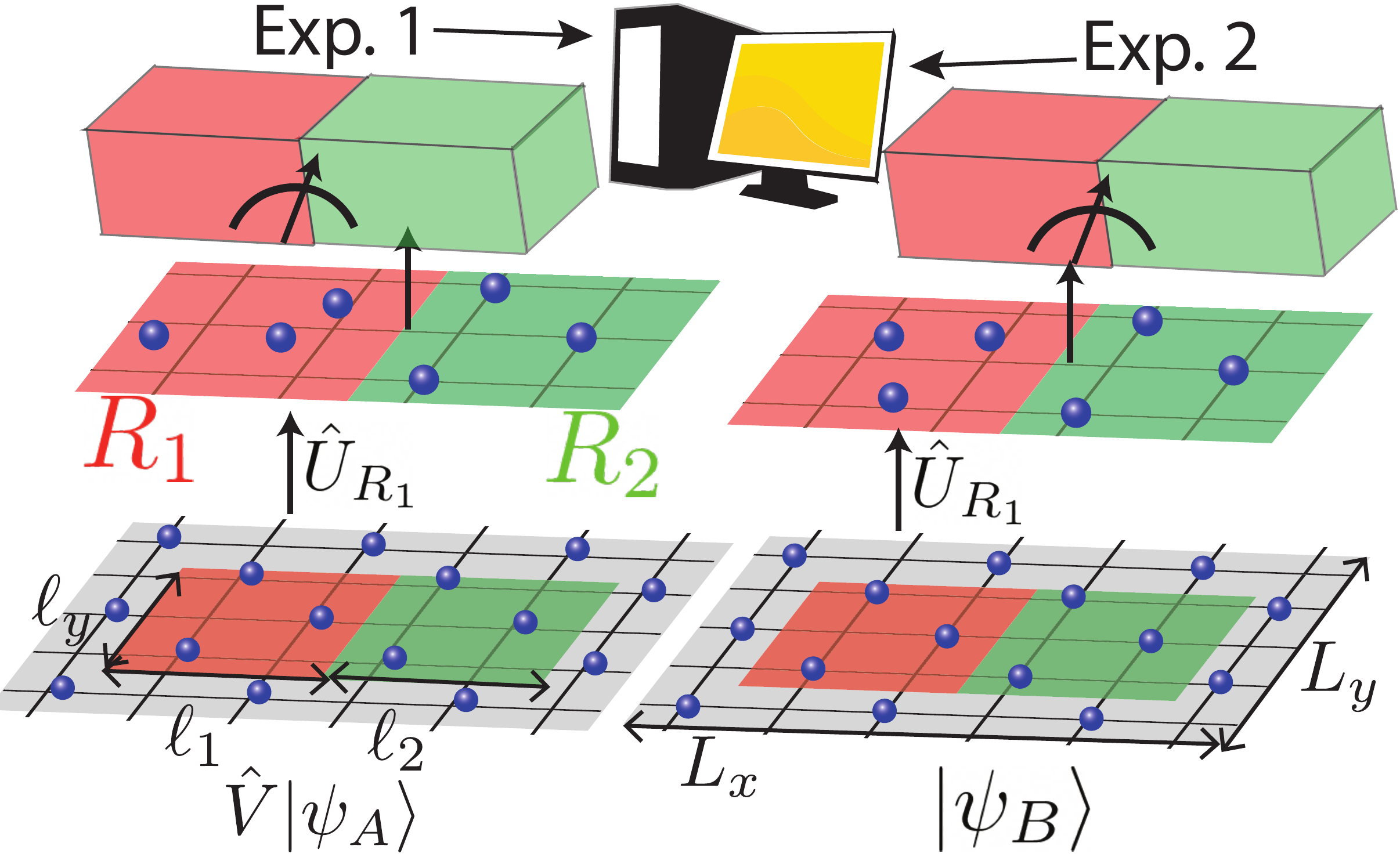}
      \caption{ The randomized measurement scheme. We define two regions $R_1$ (red) and $R_2$ (green) in the lattice with side length $\ell_1 \times \ell_y$ and $\ell_2 \times \ell_y$ respectively. We prepare two identical wave functions $\ket{\psi_A}$ and $\ket{\psi_B}$ in experiment $A$ and $B$ respectively. The local unitary operator $\hat{V}$ is applied in the region $R_1$ in the exp. $1$. Subsequently, the random unitary $\hat{U}_{R_1}$ is applied in the region $R_1$ on both wave functions. The projective measurements on the particle occupation basis are performed on regions $R_1$ and $R_2$ in both experiments. The MBCN can be inferred from the statistical correlation between the randomized measurement results in experiment $A$ and experiment $B$. }
      \label{fig_1}
\end{figure}
 
Here, we propose a novel method for the measurement of MBCN. Using our recent findings \cite{ourprb}, we show that given a wave function on a disk-like geometry, for a single set of parameters, one can construct the MBCN by applying certain operators on the wave function, without knowledge of the Hamiltonian. This should be contrasted with the common situation where one requires a family of many-body wave functions, e.g., different twist angles on a torus. Importantly, such a construction allows one to perform the measurements using random unitaries \citep{PhysRevA.99.052323, PhysRevA.97.023604, huang2020predicting}. Our scheme requires only a single wave function at a given time, for the same set of parameters, as schematically shown in Fig.~\ref{fig_1}. In other words, in each experimental realization, one requires only a single copy of the system, and simultaneous access to several identical copies of the wave function is not required. Therefore, this scheme can be easily implemented with the state of the art ultracold atoms, Rydberg arrays and circuit-QED platforms.
 
First, in the context of topological quantum field theory (TQFT) \cite{Witten1989}, we interpret and generalize the polarization formula for the MBCN \cite{ourprb}. Our approach is extensively discussed in Ref. \cite{ourprb}, here we outline the key concepts and results. Specifically, we demonstrate that by introducing two symmetry defects, in the space-time manifold, one can evaluate the MBCN, as an expectation value of symmetry defect operators. This allows us to effectively change the boundary conditions of the wave function. Then, by cutting and gluing space-time manifolds, we show that topologically non-trivial space-time manifolds, such as a torus, can be obtained from a given wave function on a rectangular geometry. Such operations can be obtained by applying a SWAP operator between two subregions \cite{Shiozaki2018many}. Similar to the Renyi entropy, where the expectation of the SWAP operator can be evaluated using a single copy of the wave function at a time, we show how such space-time surgery can be implemented in an experimental setting. Importantly, we show that the symmetry defects can be implemented by  post-processing the data. 

As a prerequisite for our protocol, we need to know the number of flux quanta that must be adiabatically inserted into a region of the system before a topologically trivial excitation is obtained \cite{ourprb}. As another feature of our protocol, we note that the amplitude of the SWAP expectation value decreases exponentially with the subregions area, in the absence of spatial symmetries. Moreover, the number of randomized measurements increases exponentially with the system size. Therefore, for both reasons, our protocol is particularly suitable for Noisy Intermediate-Scale Quantum (NISQ) devices \cite{preskill2018quantum}.
 
\textit{Many-Body Chern Number.---} In order to introduce the MBCN, we first consider a full multiplet of $s$ topologically degenerate ground states on a torus. The wave functions are $\Psi_\alpha(\phi_x, \phi_y)$ defined on a torus geometry, with length $L_x$ and $L_y$ along the $x$ and $y$ directions, respectively. Here $\alpha=1,\ldots ,s$ and we consider abelian quantum Hall states with Hall conductance $\sigma_{xy} = \frac{e^2}{h}\frac{p}{q}$, where $p$ and $q$ are co-prime integers and the parameter $s = q$. In this case, the parameter $s$ is the number of flux quanta that has to be inserted before a topologically trivial excitation is obtained. We note that in general, the parameter $s$ can be different from $q$ when the degenerate ground state subspace is composed of multiple topological sectors.\footnote{We define a topological sector to consist of all the degenerate ground states which can be related to each other under the operation of quantized flux insertion in the $x$ and $y$ cycles of a torus \cite{ourprb}.}.

The twisted boundary conditions are defined as $\hat{t}_j(L_k \hat{k}) \Psi(\phi_x, \phi_y)= e^{i\phi_k}\Psi(\phi_x, \phi_y) $, where $k= x, y$ and $\hat{t}_j(\vec{r})$ being the magnetic translation operator of the $j$th particle along the direction $\vec{r}$. The MBCN of a FQH system is of the form \citep{PhysRevB.31.3372}
\begin{eqnarray}
C = \frac{1}{2\pi i} \int_0^{2\pi s} d\phi_x \int_0^{2\pi }  d\phi_y \mathcal{F}(\phi_x, \phi_y), \label{eq:Chern}
\end{eqnarray}
where $\mathcal{F}(\phi_x, \phi_y) = \langle \partial_{\phi_x}\Psi_\alpha| \partial_{\phi_y}\Psi_\alpha\rangle -  \langle \partial_{\phi_y}\Psi_\alpha| \partial_{\phi_x}\Psi_\alpha\rangle $ is the Berry curvature obtained from adiabatically varying the twist angle boundary conditions ($\phi_x$,$\phi_y$), for a single wave function $|\Psi_\alpha\rangle$.

Alternatively, one can obtain the MBCN, when the wave function is given only as a function of one twist angle. Specifically, let $|\Psi_\alpha(\theta_x)\rangle$ be the ground state wave function in the presence of a flux through the $x$ direction $\oint dx A_x = \theta_x $, and we take the flux in the $y$ direction to be zero, $\oint dy A_y = 0$. We note that for the following argument, one can also consider a cylinder instead of a torus. Following Resta \citep{PhysRevLett.80.1800}, we define the polarization operator as $ R_y = \prod_{x,y} e^{i \frac{2 \pi y}{\ell_y} \hat{n}(x,y)},$ where the product is taken over the whole system. We then compute 
\begin{align}
  \label{Polarization}
\mathcal{T}(\theta_x,s) =\langle \Psi(\theta_x) | R_y^s |\Psi(\theta_x) \rangle.
\end{align}
Adiabatically changing $\theta_x$ is equivalent to applying an electric field $E_x$, which induces a current in the $y$ direction due to the Hall conductivity, which corresponds to a changing polarization along the $\hat{y}$ direction. The MBCN therefore can be obtained as 
\begin{align}
C = \frac{d}{d\theta_x} \text{arg} \mathcal{T}(\theta_x,s). 
\label{eq:Chern_diff}
\end{align}

We note that equation above converges to the MBCN in the thermodynamic limit. For systems with finite size, a more robust result can be obtained by averaging over the twist angle:  $C = \frac{1}{2\pi} \oint d \theta_x \frac{d}{d\theta_x} \text{arg } \mathcal{T}(\theta_x,s). $ The Hall conductivity corresponds to $\sigma_{H} = \frac{C}{s} \frac{e^2}{h}$. 

We note Eq.~\eqref{eq:Chern} and Eq.~\eqref{Polarization} are equivalent to each other and require toridal and cylindrical geometries, respectively. While there are theoretical proposals to implement such geometries \cite{kim2018optical, lkacki2016quantum}, an experimental realization remains challenging.

\begin{figure}[ht]
    \includegraphics[width=0.47\textwidth]{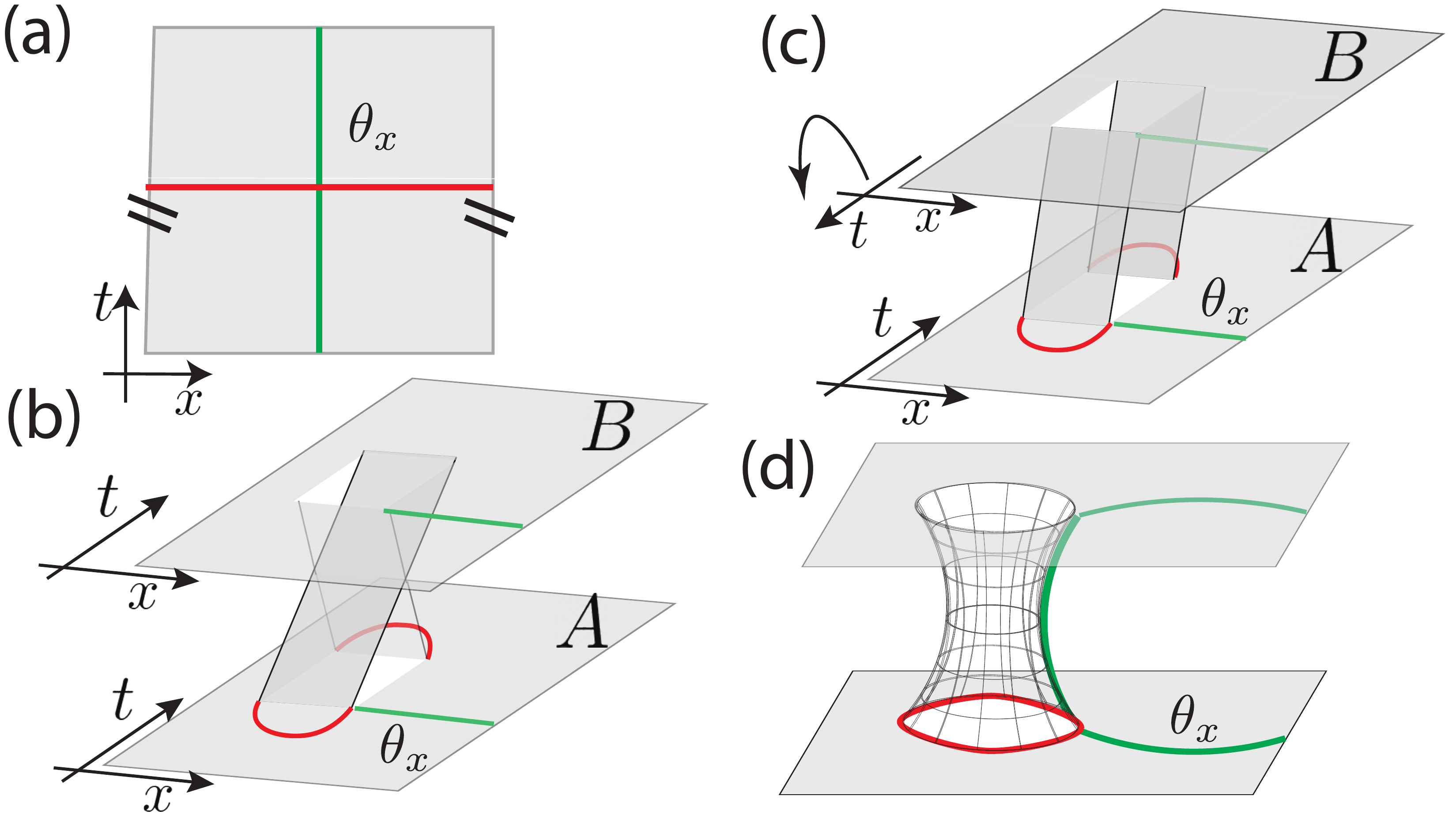}
      \caption{(a) The space-time manifold of the $\mathcal{Z}(M,A)$ in Eq. \eqref{responseTheory}, without showing the $y$ axis. The green line represents the symmetry defects $A_x$ and the red line corresponds to $A_t$. (b) The SWAP operator $\hat{S}_{R_1}$ creates a branch cut in the region $R_1$ that connects the space time between system A and system B. The red and the green curves depict the operator $\hat{V}$ and $\hat{W}(\phi)$ respectively. (c) A $\pi$ rotation around the $x$ axis in the system B maps the branch cut in (b) to a space time cylinder which is topologically equivalent to (d). }
      \label{fig_2}
\end{figure}

\textit{TQFT generalization of Resta Formula.---} We interpret and generalize the polarization formula \eqref{Polarization} using the TQFT formalism and the Chern-Simons response theory. The low-energy response of the system can be encoded in an effective action for the background electromagnetic gauge field $A$, such that the TQFT partition function on a space-time manifold $M$ is given by, 
\begin{align}
\label{responseTheory}
\mathcal{Z}(M,A) = \mathcal{Z}(M,0) e^{i \frac{p}{q} S_{CS}[A]}.
\end{align}
The Chern-Simons response action is given by $S_{CS}[A] = \frac{1}{4\pi} \int_M \epsilon^{\mu\nu\lambda} A_\mu \partial_\nu A_\lambda$, where $\mu ={t, x, y}$. The space-time manifold $M$ is $S^2\times S^1$, where $y$ and $t$ are on the sphere $S^2$ and $x$ is on the circle $S^1$. Note that $x-y$ plane forms a torus. The twisted boundary condition required in the wave functions  of Eq.\eqref{Polarization} can be realized by applying $A_{x} = \theta_x \delta(x)$ and $A_{y} = 0$. We interpret Resta's polarization operator as an application of an electric field along the $y$ direction at $t = 0$ and therefore $A_t = \frac{2\pi s y}{\ell_y}\delta (t)$. Under these conditions, the partition function is given by $\mathcal{Z}(M, A)=\mathcal{Z}(M, 0) e^{i C\theta_x}$, where $C=sp/q=p$. The background gauge fields in Eq.~\eqref{responseTheory} form two symmetry defects which are wrapped around two distinct non-contractible loops on the manifold $M$, as shown in Fig.~\ref{fig_2}(a).

Now instead of measuring the MBCN on the $x-y$ torus, here we cut and glue the space-time manifold in TQFT to construct the partition function on a topologically non-trivial manifold, by starting with the state on simple space manifolds. This allows us to create two non-contractible loops on a disk geometry. We start from two identical wave functions $\ket{\psi_A}\ket{\psi_B}$. We apply the SWAP operation $\hat{\mathbb{S}}_{R^A_{1}, R^B_1}$ between the two wave functions in the region $R_1$ as shown in Fig.~\ref{fig_1}. For an infinitesimal time interval  $\epsilon$, the SWAP operation glues the space-time manifold from $t = \mp \epsilon$ in A to $t = \pm \epsilon$ in  B, respectively, as shown in Fig.\ref{fig_2}(b). If we perform a $\pi$-rotation on the manifold of B along the $\hat{x}$ axis, it becomes clear that the two required non-contractible loops are formed, as shown in the Fig. \ref{fig_2} (c) and (d). These non-contractible loops are used to apply the symmetry defects of the gauge potential $A_t$ and $A_x$ in this synthetic non-trivial topology. 

Now, we make a connection between the TQFT and the microscopic theory to explicitly express the symmetry defects in Fig.~\ref{fig_2} in terms of the system operators. These symmetry defects are local in time and can be simply constructed by the local density operator $\hat{n}(x,y)$. Specifically, the operators that represent the polarization and the twist angle are
\begin{equation}
    \hat{V}_R = \prod_{(x,y) \in R}e^{i\frac{2\pi s y}{\ell_y} \hat{n}(x,y)}, ~
    \hat{W}_R(\theta_x) = \prod_{(x,y) \in R}e^{i\hat{n}(x,y)\theta_x}. \label{eq:symm_defect}
\end{equation}
Now the MBCN can be obtained as the expectation value of the SWAP operator, which constructs the non-trivial space-time, and the above operators. Specifically, 
\begin{equation}
\mathcal{T}(\theta_x) = \bra{\psi_A}\bra{\psi_B} \hat{V}_{R_1^A}^\dagger \hat{W}_{R_2^B}(\theta_x) \hat{\mathbb{S}}_{R^A_{1}, R^B_1} \hat{W}^\dagger_{R_2^A}(\theta_x)\hat{V}_{R_1^A}\ket{\psi_A}\ket{\psi_B},
\label{SWAP}
\end{equation}
where $R^{A(B)}_{i}$ is the $i$th region of the wave function $\ket{\psi_{A(B)}}$, $\hat{\mathbb{S}}_{R^A_{1}, R^B_1}$ is the swap operation between the two copy of the wave function and $\mathcal{T}(\theta_x) \propto e^{iC\theta_x}$. Therefore, the winding number of ${\rm arg}[\mathcal{T}(\theta_x)]$ corresponds to the MBCN. We note that while our TQFT derivation of this formula is applicable to cylindrical geometries, extensive numerical simulations indicates that the same formula can also be applied to disk-like geometries \cite{ourprb}.

\paragraph{Randomized Measurement Scheme.---} We now present the experimental protocol to measure the MBCN via random measurements. Eq. \eqref{SWAP} involves the SWAP operator between two copies of the wave function, and the expectation value can be obtained by performing a beam-splitter interaction between the two copies and a parity measurement \citep{PhysRevLett.109.020505,PhysRevLett.109.020504,islam2015measuring,kaufman2016quantum}. In contrast, we show that a random measurement protocol requires only a single wave function, at a given time. Our key observation is that, without the symmetry defect operators, Eq.\eqref{SWAP} is reminiscent of the second Renyi entropy expression and its evaluation through the SWAP operator expectation value, which can be extracted using randomized measurement \citep{PhysRevA.99.052323}. Here, we need to generalize that scheme to incorporate the symmetry defect operators.

Let us consider a two-dimensional square lattice system with open boundary condition. Eq.~\eqref{SWAP} involves non-local SWAP operations between two replica of the wave functions. It can be performed through the following two randomized measurements as described in Fig.~\ref{fig_1}. 

We start by preparing the wave function $\ket{\psi}$ in the open boundary condition. We first apply the operator $\hat{V}_{R_1}$ on the state in the experiment A. We then perform the random unitary operation $\hat{U}$ and the measurements on the occupation probability in the region $R_1$ and $R_2$ for both experiment $A$ and $B$. The random unitary operations are sampled from an approximate unitary 2-design \citep{collins2006integration,puchala2011symbolic}. After repeating the measurement $N_M$ times, we obtain the probability distribution over the occupation basis $\ket{b}$. The results of the two experiments are $P^V_{U}(b) = |\langle b| \hat{U} \hat{V}| \psi \rangle|^2$ and $P_{U}(b') =  |\langle b'| \hat{U} | \psi \rangle|^2$ respectively. We repeat the two experiments with different random unitary operations $\hat{U}$ for $N_U$ times. The statistical correlation of the measurement results in the experiment $A$ and $B$ gives

\begin{align}
\tilde{\mathcal{T}}(\theta_x) = \sum_{\{b\}}\sum_{\{b'\}}O_{b,b'}(\theta_x)\overline{P^V_{U}(b) P_{U}(b') },
\label{randomized measurement}
\end{align}
where the bar, $\overline{\cdots}$, means the average over the random unitaries from an approximate unitary 2-design. The coefficient $O_{b,b'}(\theta_x)  = \delta_{N_{1}(b), N_{1}(b')}\mathcal{D}_b(-\mathcal{D}_b)^{\delta_{b,b'}-1} e^{i[N_{2}(b)-N_{2}(b')]\theta_x}$, where $N_{1}(b)$ and $N_{2}(b)$ are the number of particles of the basis state $|b\rangle$ in the region $R_1$ and $R_2$ respectively and $\mathcal{D}_b = \binom{\ell_1\ell_y}{N_1(b)}$. Since $\tilde{\mathcal{T}}(\theta_x) = \mathcal{T}(\theta_x)$ for an ensemble average over a unitary 2-design \footnote{See supplementary for the derivation.}, the winding number of the measurement result ${\rm arg}[\tilde{\mathcal{T}}(\theta_x)]$ gives the Chern number $\tilde{C}$. 

In the following, we consider the randomized measurement scheme for system with non-trivial Chern number with finite number of $N_U$ and number of projective measurements $N_M$ for each realization of randomized measurement.

\textit{Numerical results. --} We present the measurement of MBCN for bosonic fractional quantum Hall states with filling $\nu = 1/2$. We consider hard-core boson on the $N_x \times N_y$ square lattice in the open boundary condition, with a magnetic tunneling Hamiltonian of the form 
\begin{align}
H_{\rm{t}} = -J\sum_{x,y} (\hat{a}^\dagger_{x+1,y} \hat{a}_{x,y} + e^{-i \Phi x}\hat{a}^\dagger_{x,y+1} \hat{a}_{x,y}) + \rm{h.c.},
\label{H_system}
\end{align}
where $\hat{a}_{x,y}(\hat{a}^\dagger_{x,y})$ is the bosonic annihilation (creation) operator on site $(x,y)$,  $\Phi = 2\pi/q$ is the magnetic flux on each plaquette. The ground state is known to be a FCI phase, with the MBCN $C = 1$ \citep{hafezi2007fractional,motruk2017phase, PhysRevB.96.201103}. The FCI ground state with the open boundary condition can be prepared via adiabatic process \citep{motruk2017phase, PhysRevB.96.201103}  \footnote{See supplementary for details} and engineered dissipation \citep{PhysRevX.4.031039}. We note that the system size of our simulation is within reach to the state of the art quantum computation platform \citep{arute2019quantum}.

\begin{figure}[h]
    \includegraphics[width=0.5\textwidth]{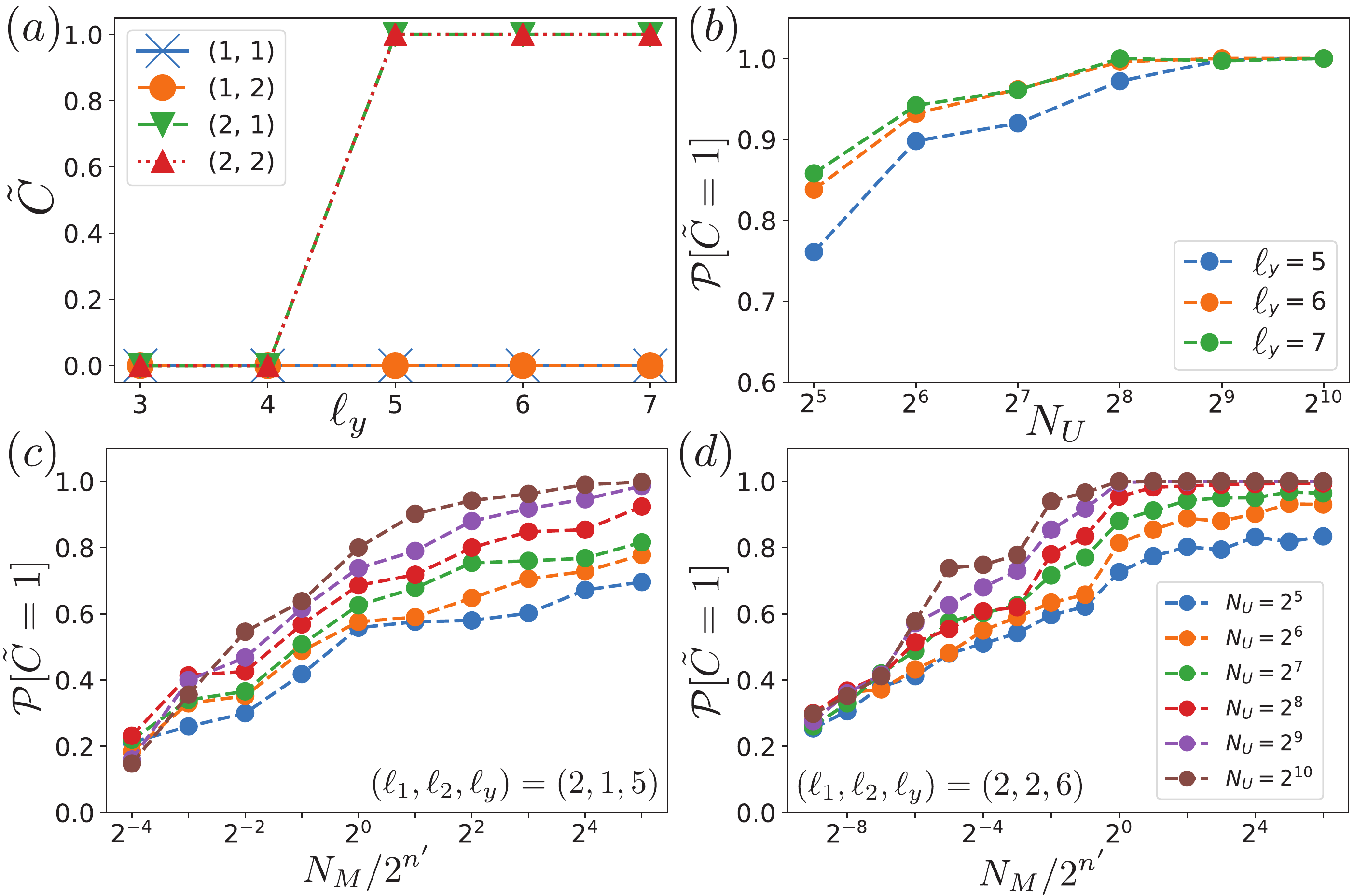}
      \caption{Simulation results for Eq. \eqref{SWAP} and \eqref{randomized measurement}, for the FCI phase with $C = 1$. (a) Obtained MBCN by Eq. \eqref{SWAP} for various region size $(\ell_1, \ell_2)$ and $\ell_y$ with $N_x = 6$, $N_y = 8$. , labeled with different markers. (b) Probability of obtaining the expected MBCN ($\mathcal{P}[\tilde{C} = 1]$) from Eq. \eqref{randomized measurement}, using randomized measurements, as a function of the number of random unitary operations $N_U$ with $N_M = \infty$. Region sizes are taken to be $\ell_1 = \ell_2 = 2$. (c, d) Probability of obtaining the expected MBCN versus number of measurements $N_M$, for two sets of region sizes. For all panels, $J = 1$, and $\Phi = 2\pi/3$. the probability $\mathcal{P}[\tilde{C}]$ is computed by averaging over 500 times independent randomized measurement results. Random quench parameters are $\eta = 20$, $\Delta =J$, $T = J^{-1}$ and $n' = 0.5n_1+n_2$. }
      \label{fig_3}
\end{figure}

In Fig.~\ref{fig_3}(a), we first show that the MBCN of this phase can be extracted, using the SWAP operator formula, Eq. \eqref{SWAP}. We observe that the correct quantized value $\tilde{C} = 1$ can be obtained,  when the region size is larger than the magnetic length of the system,  which is less than a lattice spacing in our case.   

Then, in Fig.~\ref{fig_3}(b-d), we show that the MBCN can be extracted using randomized measurement (Eq.\eqref{randomized measurement}). In order to implement random unitaries, we apply quench dynamics \citep{PhysRevA.97.023604}. We consider the number conserving random quench unitary operation $\hat{U} = \prod_{k = 1}^\eta e^{-iH_{q_k} T}$, where $\eta$ is the depth of the random quench, $T$ is the time step of each quench. The $k$th quench Hamiltonian is of the form
\begin{align}
&H_{q_k} = -J \sum_{\langle i,j\rangle,~i, j \in R_1} (a^\dagger_{i}a_{j} +\rm{h.c.}) + \sum_{i \in R_1} \Delta^k_i \hat{n}_i,
\end{align}
where $\Delta^k_i$ is a Gaussian distributed random number with mean zero and standard deviation $\Delta$. It has been shown that when the magnitude of $\Delta$ is comparable to $T^{-1}$ and $J$, the random quench unitary operator gives the approximate 2-design unitary  \citep{PhysRevA.97.023604}.

The performance of the randomized measurement is characterized by the probability of obtaining the correct MBCN $\mathcal{P}[\tilde{C} = 1]$. In Fig. \ref{fig_3}(b), we consider the limit of $N_M \rightarrow \infty$, the performance of the randomized measurement weakly depends on the number of qubits in the measurement region $R_1$ and $R_2$. In Fig. \ref{fig_3} (c) and (d), the shot-noise of the measurements are taken into account. When the number of measurements $N_M$ is of the same order of magnitude as $2^{n'}$, where $n' = 0.5n_1 + n_2$, and $n_1$ and $n_2$ are the number of sites in the region $R_1$ and $R_2$ respectively, the probability $\mathcal{P}[\tilde{C} = 1]$ starts to saturate. The factor $2^{0.5n_1}$ originates from the birthday paradox scaling of the randomized measurement in the region $R_1$ \citep{PhysRevA.99.052323} and the factor $2^{n_2}$ is contributed by the shot-noise of the number operator measurement in the region $R_2$. The randomized measurements can be realized in the current and near-term experimental platform. For example, in the circuit QED architecture with 10kHz repetition rate, each randomized measurement can be performed within a few minutes.

\paragraph{Robustness against errors of the NISQ devices.}
In order to demonstrate the feasibility in the NISQ devices, we show that our protocol is robust against various types of experimental imperfections. First, we note that the randomize measurement protocol is robust against the small miscalibration of the quantum hardware. It has been shown that the leading order contribution of the miscalibration vanish in the randomized measurement protocol \cite{PhysRevX.9.021061}.  

 For the amplitude damping error and the readout error, since the total number of excitations in the whole system is conserved during the state preparation and random unitary gate, when either the amplitude damping error or the readout error occurs, the total number of excitation is changed. A change of the number of excitation heralds an error and the run should be discarded. Therefore, up to the first order of the error rates, the amplitude damping error or the readout error can be detected.

In the case of the depolarization error, the quantum state after performing the random unitary operation is of the form
\begin{eqnarray}
\rho_{\rm dep} = (1-p_{\rm dep})\rho_{\rm ideal} + \frac{p_{\rm dep}}{\mathcal{D}} I_{\mathcal{D}} + O(p_{\rm dep}^2),
\end{eqnarray}
where $\rho_{\rm ideal}$ is the density matrix in the ideal situation, $\mathcal{D}$ is the dimension of the Hilbert space and $p_{\rm dep}$ is the depolarization probability. After performing the measurement and post-processing described in Eq. \eqref{randomized measurement}, we have
\begin{eqnarray}
\tilde{\mathcal{T}}_{\rm dep}(\theta_x) \approx (1-p_{\rm dep})^2\tilde{\mathcal{T}}(\theta_x) + p_{\rm dep}c(\theta_x),
\label{T_dephasing}
\end{eqnarray}
where $c(\theta_x)$ is a constant offset which can be calculated from the measurement results \footnote{See supplementary for the derivation.}. 

Since the amplitude of $\mathcal{\tilde{T}}(\theta_x)$ is rescaled, the number of measurements should be increased in order to have converged results. For example, one can increase the number of random unitary $N_U$ by a factor of $\frac{1}{(1-p_{\rm dep})^4}$ in order to increase the measurement accuracy by a factor of $(1-p_{\rm dep})^2$ \citep{PhysRevA.99.052323}. The winding number of $\tilde{\mathcal{T}}(\theta_x)$ can be extracted by fitting the measurement result in Eq. \eqref{T_dephasing} with parameters $\theta_x$ and $p_{\rm dep}$. 

\textit{Outlook.--}
Our work opens up a new avenue for creating non-trivial topology on space-time manifold, using the SWAP operation. It is particularly intriguing that the SWAP operation can be implemented by random unitaries in the NISQ devices. More broadly, quantum simulators are poised to realize topologically-ordered states that might not occur in a conventional electronic matter. Given this opportunity, it is important to develop measurement methods that go beyond linear response formalism. For example, it is interesting to investigate whether the application of SWAP operator through randomized measurement can be used to probe other topological characterizations, such as modular matrices \cite{zhu2017}, topological entanglement entropy \cite{kitaev2006topological,levin2006detecting} 
and the order parameter of the symmetry enriched topological phases \citep{garre2019local}.

\textit{Acknowledgments.--} ZC thanks Hsin-Yuan Huang for helpful discussion about randomized measurement. BV thanks C. Repellin for discussions. ZC, HD, and MH were supported by AFOSR FA9550-16-1-0323, FA9550- 19-1-0399, ARO W911NF-15-1-0397 and Google AI. HD, MH thank the hospitality of the Kavli Institute for Theoretical Physics, supported by NSF PHY-1748958. MB is supported by NSF CAREER (DMR-1753240), Alfred P. Sloan Research Fellowship. ZC, HD, MH, and MB acknowledge the support of NSF Physics Frontier Center at the Joint Quantum Institute. AE, BV and PZ were supported by the European Union's Horizon 2020 research and innovation programme under Grant Agreement No. 817482 (PASQuanS) and No. 731473 (QuantERA via QTFLAG) and by the Simons Collaboration on Ultra-Quantum Matter, which is a grant from the Simons Foundation (651440, P. Z.). BV acknowledges funding from the Austrian Science Fundation (FWF) with the project P-32597.

\bibliographystyle{unsrtnat}
\bibliography{topobib.bib}

\end{document}